\input harvmac
%\draftmode
\let\includefigures=\iftrue
\let\useblackboard==\iftrue
\newfam\black

%Figure Stuff
\includefigures
\message{If you do not have epsf.tex (to include figures),}
\message{change the option at the top of the tex file.}
\input epsf
\def\figin{\epsfcheck\figin}\def\figins{\epsfcheck\figins}
\def\epsfcheck{\ifx\epsfbox\UnDeFiNeD
\message{(NO epsf.tex, FIGURES WILL BE IGNORED)}
\gdef\figin##1{\vskip2in}\gdef\figins##1{\hskip.5in}% blank space instead
\else\message{(FIGURES WILL BE INCLUDED)}%
\gdef\figin##1{##1}\gdef\figins##1{##1}\fi}
\def\DefWarn#1{}
\def\figinsert{\goodbreak\midinsert}
\def\ifig#1#2#3{\DefWarn#1\xdef#1{fig.~\the\figno}
\writedef{#1\leftbracket fig.\noexpand~\the\figno}%
\figinsert\figin{\centerline{#3}}\medskip\centerline{\vbox{
\baselineskip12pt\advance\hsize by -1truein
\noindent\footnotefont{\bf Fig.~\the\figno:} #2}}
%\bigskip
\endinsert\global\advance\figno by1}
%%%
\else
\def\ifig#1#2#3{\xdef#1{fig.~\the\figno}
\writedef{#1\leftbracket fig.\noexpand~\the\figno}%
%\figinsert\figin{\centerline{#3}}\medskip
%\centerline{\vbox{\baselineskip12pt
%\advance\hsize by -1truein\noindent
%\footnotefont{\bf Fig.~\the\figno:} #2}}
%\bigskip\endinsert
\global\advance\figno by1} \fi

\def\id{{1 \kern-.28em {\rm l}}}

\def\K3{{\bf K3}}
\def\journal#1&#2(#3){\unskip, \sl #1\ \bf #2 \rm(19#3) }
\def\andjournal#1&#2(#3){\sl #1~\bf #2 \rm (19#3) }

\def\bar{\overline}
\def\hat{\widehat}
\def\ie{{\it i.e.}}
\def\eg{{\it e.g.}}

\def\frac#1#2{{#1\over#2}}

\def\half{\frac12}

\def\inbar{\,\vrule height1.5ex width.4pt depth0pt}
\def\IC{\relax\hbox{$\inbar\kern-.3em{\rm C}$}}
\def\IR{\relax{\rm I\kern-.18em R}}
\def\IP{\relax{\rm I\kern-.18em P}}

%
%%%%%%%%%%%%%%%%%%%%%%%%%%%%%%%%%%%%
%

%
\catcode`\@=11
\def\slash#1{\mathord{\mathpalette\c@ncel{#1}}}
\overfullrule=0pt

\def\LL{{\cal L}}
\def\MM{{\cal M}}
\def\NN{{\cal N}}
\def\OO{{\cal O}}

\def\underrel#1\over#2{\mathrel{\mathop{\kern\z@#1}\limits_{#2}}}

\catcode`\@=12

%%%%%%%%%%%%%%%%%%%%%%%%%%%%%%%%%%%%%%%%%%%%%%%%%%%%%%%%%%%%%%

%

%%%%%%%%%%%%%%%%%%%%%%%%%%%%%%%%%%%%%%%%%%%%%%%%%%%%%%%%%%%%%%
% new defs:

\def\ie{{\it i.e.}}
\def\eg{{\it e.g.}}

%%%%%%%%%%%%%%%%%%%%%%%%%%%%%%%%%%%%%%%%%%%%%%%%%%
%\IsraelIR
\lref\IsraelIR{
  D.~Israel, C.~Kounnas, A.~Pakman and J.~Troost,
  ``The Partition function of the supersymmetric two-dimensional black hole and little string theory,''
JHEP {\bf 0406}, 033 (2004).
[hep-th/0403237].
%%CITATION = hep-th/0403237%%
}

%\MaldacenaHW
\lref\MaldacenaHW{
  J.~M.~Maldacena and H.~Ooguri,
  ``Strings in AdS(3) and SL(2,R) WZW model 1.: The Spectrum,''
J.\ Math.\ Phys.\  {\bf 42}, 2929 (2001).
[hep-th/0001053].
%%CITATION = hep-th/0001053%%
}

%\ZamolodchikovCE
\lref\ZamolodchikovCE{
  A.~B.~Zamolodchikov,
  ``Expectation value of composite field T anti-T in two-dimensional quantum field theory,''
[hep-th/0401146].
%%CITATION = hep-th/0401146%%
}

%\SmirnovLQW
\lref\SmirnovLQW{
  F.~A.~Smirnov and A.~B.~Zamolodchikov,
  ``On space of integrable quantum field theories,''
[arXiv:1608.05499 [hep-th]].
%%CITATION = arXiv:1608.05499%%
}

%\CavagliaODA
\lref\CavagliaODA{
  A.~Cavagliˆ, S.~Negro, I.~M.~SzŽcsŽnyi and R.~Tateo,
  ``$T \bar{T}$-deformed 2D Quantum Field Theories,''
[arXiv:1608.05534 [hep-th]].
%%CITAT
}

%\GiveonNIE
\lref\GiveonNIE{
  A.~Giveon, N.~Itzhaki and D.~Kutasov,
  ``$T\bar T$ and LST,''
[arXiv:1701.05576 [hep-th]].
%%CITATION = arXiv:1701.05576%%
}

%\PolchinskiRQ
\lref\PolchinskiRQ{
  J.~Polchinski,
  ``String theory. Vol. 1: An introduction to the bosonic string,''
}

%\GiveonCGS
\lref\GiveonCGS{
  A.~Giveon and D.~Kutasov,
  ``Supersymmetric Renyi entropy in CFT$_{2}$ and AdS$_{3}$,''
JHEP {\bf 1601}, 042 (2016).
[arXiv:1510.08872 [hep-th]].
%%CITATION = arXiv:1510.08872%%
}

%\KutasovXU
\lref\KutasovXU{
  D.~Kutasov and N.~Seiberg,
  ``More comments on string theory on AdS(3),''
JHEP {\bf 9904}, 008 (1999).
[hep-th/9903219].
%%CITATION = hep-th/9903219%%
}

%\GiveonNS
\lref\GiveonNS{
  A.~Giveon, D.~Kutasov and N.~Seiberg,
  ``Comments on string theory on AdS(3),''
Adv.\ Theor.\ Math.\ Phys.\  {\bf 2}, 733 (1998).
[hep-th/9806194].
%%CITATION = hep-th/9806194%%
}

%\GiveonUP
\lref\GiveonUP{
  A.~Giveon and D.~Kutasov,
  ``Notes on AdS(3),''
Nucl.\ Phys.\ B {\bf 621}, 303 (2002).
[hep-th/0106004].
%%CITATION = RI-5-01%%
}

%\MaldacenaKM
\lref\MaldacenaKM{
  J.~M.~Maldacena and H.~Ooguri,
  ``Strings in AdS(3) and the SL(2,R) WZW model. Part 3. Correlation functions,''
Phys.\ Rev.\ D {\bf 65}, 106006 (2002).
[hep-th/0111180].
%%CITATION = CALT-68-2360%%
}

%\ArgurioTB
\lref\ArgurioTB{
  R.~Argurio, A.~Giveon and A.~Shomer,
  ``Superstrings on AdS(3) and symmetric products,''
JHEP {\bf 0012}, 003 (2000).
[hep-th/0009242].
%%CITATION = hep-th/0009242%%
}

%\GiveonMI
\lref\GiveonMI{
  A.~Giveon, D.~Kutasov, E.~Rabinovici and A.~Sever,
  ``Phases of quantum gravity in AdS(3) and linear dilaton backgrounds,''
Nucl.\ Phys.\ B {\bf 719}, 3 (2005).
[hep-th/0503121].
%%CITATION = hep-th/0503121%%
}

%\GiveonJG
\lref\GiveonJG{
  A.~Giveon and M.~Rocek,
  ``Supersymmetric string vacua on AdS(3) x N,''
JHEP {\bf 9904}, 019 (1999).
[hep-th/9904024].
%%CITATION = hep-th/9904024%%
}

%\BerensteinGJ
\lref\BerensteinGJ{
  D.~Berenstein and R.~G.~Leigh,
  ``Space-time supersymmetry in AdS(3) backgrounds,''
Phys.\ Lett.\ B {\bf 458}, 297 (1999).
[hep-th/9904040].
%%CITATION = hep-th/9904040%%
}

\lref\gik{Work in progress.}

%\GiveonZM
\lref\GiveonZM{
  A.~Giveon, D.~Kutasov and O.~Pelc,
  ``Holography for noncritical superstrings,''
JHEP {\bf 9910}, 035 (1999).
[hep-th/9907178].
%%CITATION = hep-th/9907178%%
}

%\IsraelRY
\lref\IsraelRY{
  D.~Israel, C.~Kounnas and M.~P.~Petropoulos,
  ``Superstrings on NS5 backgrounds, deformed AdS(3) and holography,''
JHEP {\bf 0310}, 028 (2003).
[hep-th/0306053].
%%CITATION = hep-th/0306053%%
}

%\ForsteWP
\lref\ForsteWP{
  S.~Forste,
  ``A Truly marginal deformation of SL(2, R) in a null direction,''
Phys.\ Lett.\ B {\bf 338}, 36 (1994).
[hep-th/9407198].
%%CITATION = hep-th/9407198%%
}

%\GiveonGB
\lref\GiveonGB{
  A.~Giveon, E.~Rabinovici and A.~Sever,
  ``Strings in singular time dependent backgrounds,''
Fortsch.\ Phys.\  {\bf 51}, 805 (2003).
[hep-th/0305137].
%%CITATION = hep-th/0305137%%
}

%\AharonyVK
\lref\AharonyVK{
  O.~Aharony, B.~Fiol, D.~Kutasov and D.~A.~Sahakyan,
  ``Little string theory and heterotic / type II duality,''
Nucl.\ Phys.\ B {\bf 679}, 3 (2004).
[hep-th/0310197].
%%CITATION = hep-th/0310197%%
}

%\BerkoozUG
\lref\BerkoozUG{
  M.~Berkooz, A.~Sever and A.~Shomer,
  ``'Double trace' deformations, boundary conditions and space-time singularities,''
JHEP {\bf 0205}, 034 (2002).
[hep-th/0112264].
%%CITATION = hep-th/0112264%%
}

%\WittenUA
\lref\WittenUA{
  E.~Witten,
  ``Multitrace operators, boundary conditions, and AdS / CFT correspondence,''
[hep-th/0112258].
%%CITATION = hep-th/0112258%%
}

%\MaldacenaUZ
\lref\MaldacenaUZ{
  J.~M.~Maldacena, J.~Michelson and A.~Strominger,
  ``Anti-de Sitter fragmentation,''
JHEP {\bf 9902}, 011 (1999).
[hep-th/9812073].
%%CITATION = hep-th/9812073%%
}

%\SeibergXZ
\lref\SeibergXZ{
  N.~Seiberg and E.~Witten,
  ``The D1 / D5 system and singular CFT,''
JHEP {\bf 9904}, 017 (1999).
[hep-th/9903224].
%%CITATION = hep-th/9903224%%
}

%\AharonyXN
\lref\AharonyXN{
  O.~Aharony, A.~Giveon and D.~Kutasov,
  ``LSZ in LST,''
Nucl.\ Phys.\ B {\bf 691}, 3 (2004).
[hep-th/0404016].
%%CITATION = hep-th/0404016%%
}

%\GiveonPX
\lref\GiveonPX{
  A.~Giveon and D.~Kutasov,
  ``Little string theory in a double scaling limit,''
JHEP {\bf 9910}, 034 (1999).
[hep-th/9909110].
%%CITATION = hep-th/9909110%%
}

%\GiveonTQ
\lref\GiveonTQ{
  A.~Giveon and D.~Kutasov,
  ``Comments on double scaled little string theory,''
JHEP {\bf 0001}, 023 (2000).
[hep-th/9911039].
%%CITATION = hep-th/9911039%%
}

%\HorneGN
\lref\HorneGN{
  J.~H.~Horne and G.~T.~Horowitz,
  ``Exact black string solutions in three-dimensions,''
Nucl.\ Phys.\ B {\bf 368}, 444 (1992).
[hep-th/9108001].
%%CITATION = hep-th/9108001%%
}

%\HorowitzEI
\lref\HorowitzEI{
  G.~T.~Horowitz and A.~A.~Tseytlin,
  ``On exact solutions and singularities in string theory,''
Phys.\ Rev.\ D {\bf 50}, 5204 (1994).
[hep-th/9406067].
%%CITATION = hep-th/9406067%%
}

%\ItzhakiZR
\lref\ItzhakiZR{
  N.~Itzhaki, D.~Kutasov and N.~Seiberg,
  ``Non-supersymmetric deformations of non-critical superstrings,''
JHEP {\bf 0512}, 035 (2005).
[hep-th/0510087].
%%CITATION = hep-th/0510087%%
}

%\MartinecZTD
\lref\MartinecZTD{
  E.~J.~Martinec and S.~Massai,
  ``String Theory of Supertubes,''
[arXiv:1705.10844 [hep-th]].
%%CITATION = arXiv:1705.10844%%
}

%\CallanAT
\lref\CallanAT{
  C.~G.~Callan, Jr., J.~A.~Harvey and A.~Strominger,
  ``Supersymmetric string solitons,''
In *Trieste 1991, Proceedings, String theory and quantum gravity '91* 208-244 and Chicago Univ. - EFI 91-066 (91/11,rec.Feb.92) 42 p.
[hep-th/9112030].
%%CITATION = hep-th/9112030%%
}

%%%%%%%%%%%%%%%%%%%%%%%%%%%%%%%%%%%%%%%%%%%%%%%%%%%
\Title{} {\centerline{A Solvable Irrelevant Deformation of $AdS_3/CFT_2$}}

\bigskip
\centerline{\it Amit Giveon${}^{1}$, Nissan Itzhaki${}^{2}$ and David Kutasov${}^{3}$}
\bigskip
\smallskip
\centerline{${}^{1}$Racah Institute of Physics, The Hebrew
University} \centerline{Jerusalem 91904, Israel}
\smallskip
\centerline{${}^{2}$ Physics Department, Tel-Aviv University, Israel} \centerline{Ramat-Aviv, 69978, Israel}
\smallskip
\centerline{${}^3$EFI and Department of Physics, University of
Chicago} \centerline{5640 S. Ellis Av., Chicago, IL 60637, USA }
\smallskip

\vglue .3cm

\bigskip

\bigskip
\noindent

Recently we proposed a universal solvable irrelevant deformation of $AdS_3/CFT_2$ duality, which leads in the ultraviolet to a theory with a Hagedorn entropy \GiveonNIE. In this note we provide a worldsheet description of this theory as a coset CFT, and compare its spectrum to the field theory predictions of \refs{\SmirnovLQW,\CavagliaODA}.

\bigskip

\Date{}

%%%%%%%%%%%%%%%%%%%%%%%%%%%%%%%%%%%%%%%%%%%%%%%%%%%%%%%%%%%%%%%%
%%%%%%%%%%%%%%%%%%%%%%%%%%%%%%%%%%%%%%%%%%%%%%%%%%%%%%%%%%%%%%%%

\newsec{Introduction}

In this note we continue our  study \GiveonNIE\ of a certain deformation of string theory on $AdS_3$. This study was motivated
by two recent papers \refs{\SmirnovLQW,\CavagliaODA}, which argued that perturbing a two dimensional conformal field theory ($CFT_2$)  by a particular dimension $(2,2)$ operator, which behaves near the original CFT like the product of the holomorphic and anti-holomorphic components of the stress tensor, $T\bar T$, leads to a well defined theory, despite the fact that it corresponds to a flow up the renormalization group (RG). Moreover, the authors of these papers argued that the model is in a certain sense exactly solvable, and in particular computed its spectrum on $\IR\times S^1$. An interesting property of the resulting spectrum is that it smoothly interpolates between an entropy associated with a $CFT_2$ in the IR, and one that exhibits Hagedorn growth in the UV \GiveonNIE.

In the context of holography, the irrelevant deformation studied in \refs{\SmirnovLQW,\CavagliaODA} is a double trace deformation, which corresponds to a change of the boundary conditions of the bulk fields on $AdS_3$ \refs{\BerkoozUG,\WittenUA}. In \GiveonNIE, we pointed out that there is a single trace deformation of string theory on $AdS_3$ that shares many elements with that of \refs{\SmirnovLQW,\CavagliaODA}, but may be more interesting, since it modifies the local geometry of the bulk theory. Some of the features the two deformations have in common are:
\item{(1)} The perturbing operator is a quasi-primary of the (boundary, or spacetime) Virasoro algebra with dimension $(2,2)$. Moreover, the OPE of the perturbing operator with the stress tensor has the same structure in the two cases.
\item{(2)} The construction of \refs{\SmirnovLQW,\CavagliaODA} is universal, in the sense that all $CFT_2$'s contain the operator $T\bar T$ that drives the RG flow. Similarly, the construction of \GiveonNIE\ is universal, in the sense that the single trace operator that drives the RG flow exists in all vacua of string theory on $AdS_3$.
\item{(3)} In the string theory construction, the irrelevant deformation of the spacetime theory corresponds to a marginal deformation of the worldsheet one. Therefore, from the string theory perspective it is natural that the resulting spacetime theory is well defined, as in \refs{\SmirnovLQW,\CavagliaODA}.
\item{(4)} The marginal worldsheet deformation is by an operator bilinear in worldsheet currents, and as such is exactly solvable, as in \refs{\SmirnovLQW,\CavagliaODA}. In fact, as mentioned in \GiveonNIE\ and will be further discussed below, the deformed worldsheet theory can be thought of as a coset CFT, and one can use current algebra techniques to study it.
\item{(5)} The string theory construction of \GiveonNIE\ gives rise to a theory that interpolates between a $CFT_2$ entropy in the IR and a Hagedorn entropy in the UV, like in \refs{\SmirnovLQW,\CavagliaODA}.

Despite the close analogy between the two constructions, the precise relation between them is unclear, primarily due to our limited understanding of the spacetime CFT corresponding to string theory on $AdS_3$. In \GiveonNIE, it was pointed out that if we assume that the spacetime CFT takes the symmetric product form $\MM^p/S_p$, where $\MM$ is a CFT with central charge $6k$, and  $k$ is the level of the worldsheet $SL(2,\IR)$ current algebra in string theory on $AdS_3$, as suggested in \refs{\ArgurioTB,\GiveonCGS},
the string theory single trace deformation corresponds to a $T\bar T$ deformation of the block $\MM$. The high energy behavior of the entropy of the deformed symmetric product CFT was shown to agree with the Bekenstein-Hawking entropy of black holes in the deformed geometry induced by the single trace deformation.

In this note, we would like to comment on a few aspects of the construction of \GiveonNIE. In section 2, we describe this construction in terms of a coset CFT, which involves null gauging of a $10+2$ dimensional background. We comment briefly on observables in the theory, which are naturally described in terms of this coset CFT, and use it (in section 3) to describe the spectrum of states of the resulting model on a spatial circle. We show that for the superstring, in a particular vacuum with supersymmetry preserving boundary conditions on the circle, the spectrum one gets is the same as that of \refs{\SmirnovLQW,\CavagliaODA}, assuming the $\MM^p/S_p$ structure mentioned above. In section 4, we discuss our results and their relation to those of \GiveonMI\ on the string/black hole transition in $AdS_3$ and linear dilaton backgrounds.

\newsec{Coset description}

A large class of $(2,2)$ supersymmetric vacua of string theory on $AdS_3$ is obtained by studying the worldsheet theory on $AdS_3\times S^1\times \NN$, where $\NN$ is a compact background described by a $(2,2)$ superconformal worldsheet theory
(see \eg\ \refs{\GiveonNS,\GiveonJG,\BerensteinGJ,\ArgurioTB}). Spacetime SUSY leads to a chiral GSO projection, which acts as an orbifold on this background. A useful way of thinking about these backgrounds is as describing systems of $NS5$-branes wrapped around various surfaces in a way that preserves some supersymmetry, in a state with a large number of fundamental strings bound to the fivebranes \refs{\GiveonZM,\ArgurioTB}.

A special case of this construction, which is sufficient for our purposes, is the background corresponding to $k$ $NS$ fivebranes wrapped around a four manifold $\MM^4(=T^4$ or $K_3)$, and $p$ strings,
\eqn\aaa{AdS_3\times S^3\times \MM^4.}
As in \GiveonNIE, we are interested in deforming this background by adding to the worldsheet Lagrangian the term
\eqn\deltall{\delta\CL=\lambda J^-\bar J^-,}
where $J^-$ is the worldsheet $SL(2,\IR)$ current whose zero mode gives rise to the spacetime Virasoro generator $L_{-1}$. As described in \refs{\ForsteWP,\GiveonZM,\IsraelRY}, this marginal worldsheet deformation leads to an asymptotically linear dilaton geometry, which interpolates between the $(AdS_3)$ near-horizon geometry of both the strings and the fivebranes in the IR, and the linear dilaton (CHS \CallanAT) geometry of just the fivebranes in the UV.

To describe the deformed CFT as a coset, we start with the following $10+2$ dimensional background:\foot{Or, in the more general class of vacua mentioned above, $\IR^{1,1}\times AdS_3\times S^1\times \NN$.}
\eqn\twelved{\IR^{1,1}\times AdS_3\times S^3\times \MM^4.}
Later, when studying states on the cylinder, we will compactify the spatial direction in $\IR^{1,1}$ on a circle. The uncompactified geometry is useful for studying off-shell correlation functions, as in \refs{\AharonyVK,\AharonyXN}.

We note in passing that the background \twelved\ plays an important role in many studies of fivebranes in string theory. For example, the system of fivebranes on a circle \refs{\GiveonPX,\GiveonTQ}, known as Double Scaled Little String Theory (DSLST), involves the coset of \twelved\ by the null current $J^3-K^3$, where $J^3$ is the timelike $U(1)$ in $AdS_3$ and $K^3$ is a CSA generator of $SU(2)$ \IsraelIR. Systems of fivebranes in motion are described by modifying the null current to involve the time translation generator \ItzhakiZR. And, recently it has been shown \MartinecZTD\ that some of the Ramond ground states of the string-fivebrane system can be described by adding the null translation generator in $\IR^{1,1}$ (more precisely $\IR\times S^1$) to the null generator $J^3-K^3$ mentioned above. Other closely related cosets give rise to black holes
(see e.g. \refs{\HorneGN,\HorowitzEI} and appendix C of \GiveonMI)
and cosmological backgrounds (see e.g. \GiveonGB\ for a review).

To describe the construction of \GiveonNIE\ as a coset CFT, we gauge the null current
\eqn\nullcur{i\partial(y-t)+\epsilon J^-,}
where $(t,y)$ are coordinates on $\IR^{1,1}$ and, again, $y$ may be compact. The current \nullcur\ is null and thus anomaly free. We can also gauge the right-moving current $i\bar\partial(y+t)+\epsilon\bar J^-$.

To understand the geometry that we get by gauging \nullcur\ and its right-moving analog in \twelved, we start with the sigma model on $AdS_3\times \IR^{1,1}$, which is described by the worldsheet Lagrangian
\eqn\wslag{ \LL=k(\partial\phi\bar\partial\phi+e^{2\phi}\bar\partial\gamma\partial\bar\gamma)+\partial x^+\bar\partial x^-.
}
The coordinates $\gamma=\gamma^1-\gamma^0$, $\bar\gamma=\gamma^1+\gamma^0$ parametrize the boundary of $AdS_3$; $x^\pm=y\pm t$ are coordinates on $\IR^{1,1}$. The symmetry we would like to mod out by is
\eqn\gaugesym{\eqalign{x^-\to &\,\,x^-+\alpha\;;\qquad\gamma\to\gamma+\epsilon\alpha,\cr
x^+\to &\,\,x^++\bar\alpha\;;\qquad\bar\gamma\to\bar\gamma+\epsilon\bar\alpha,
}}
where $\alpha$, $\bar\alpha$ are the gauge parameters of the two null $U(1)$'s.\foot{In \gaugesym\ we chose an axial gauging. One could also perform a vector gauging, for which  $\bar\gamma\to\bar\gamma-\epsilon\bar\alpha$. This gives rise to a singular geometry \GiveonNIE.}

To implement the gauging, we modify  \wslag\ as follows:
\eqn\gaugedlag{ \LL=k\left[\partial\phi\bar\partial\phi+e^{2\phi}(\bar\partial\gamma+\epsilon\bar A)(\partial\bar\gamma+\epsilon A)\right]+(\partial x^++A)(\bar\partial x^-+\bar A).
}
Eliminating the gauge fields gives rise to the background
\eqn\wwssll{\LL=k\partial\phi\bar\partial\phi+{k\over k\epsilon^2+e^{-2\phi}}\bar\partial(\gamma-\epsilon x^-)\partial(\bar\gamma-\epsilon x^+),
}
with a dilaton that goes like $\Phi\sim-\ln(1+k\epsilon^2e^{2\phi})$. The metric, $B$ field and dilaton depend on the gauge invariant coordinates $\phi$, $\gamma^0-\epsilon t$ and $\gamma^1-\epsilon y$. We can fix the gauge $x^\pm=0$, which is natural in the infrared region $\phi\to-\infty$, or $\gamma=\bar\gamma=0$, which is natural near the boundary $\phi\to+\infty$. This gives rise to the well known geometry of strings and fivebranes (see \eg\ appendix A of \GiveonZM).

The parameter $\epsilon$ in \wwssll\ controls the transition from the near-horizon region of both the strings and the fivebranes $(e^{-\phi}\gg\epsilon\sqrt k)$, and the region where we are in the near horizon of the fivebranes but not of the strings $(e^{-\phi}\ll\epsilon\sqrt k)$. We can set it to any particular value by shifting $\phi$ and rescaling $(\gamma,\bar\gamma)$. The role of this parameter in the bulk theory is very similar to that of the coefficient of the irrelevant operator  in the Lagrangian of the corresponding boundary theory. The latter determines the scale at which the theory transitions from being dominated by the IR CFT, and the UV (Hagedorn) regime.

As mentioned above, the coset perspective is useful for studying correlation functions of off-shell operators in the theory. We will postpone a detailed discussion of these correlation functions to another publication, limiting our discussion here to a few comments.

Setting the deformation parameter $\lambda$ in \deltall\ to zero (or, equivalently, setting $\epsilon=0$ in the coset \nullcur), off-shell observables correspond to local operators on the boundary of $AdS_3$. A large class of such observables is given by vertex operators in the (NS,NS) sector, which take the form (in the $(-1,-1)$ picture)
\eqn\locobs{\hat\OO(x)=\int d^2z e^{-\varphi-\bar\varphi}\Phi_h(x;z)\OO(z).}
Here $\varphi$, $\bar\varphi$ are worldsheet fields associated with the superconformal ghosts, that keep track of the picture. $\Phi_h(x;z)$ are natural vertex operators on $AdS_3$, labeled by position on the boundary, $x$, and on the worldsheet, $z$ (see \eg\ \refs{\KutasovXU,\GiveonUP} for more detailed discussions and references), and $\OO$ is an ($\NN=1$ superconformal primary) operator in the worldsheet theory on $S^3\times \MM^4$, or more generally $S^1\times \NN$. The operator \locobs\ satisfies the mass-shell condition
\eqn\massshell{-{h(h-1)\over k}+\Delta_\OO=\half~,}
which relates the scaling dimension of the operator $\hat\OO(x)$ in the spacetime (or boundary) CFT, $h$, to the worldsheet scaling dimension of the operator $\OO$, $\Delta_\OO$.

When we add to the theory the $\IR^{1,1}$ factor in  \twelved\ and gauge the symmetry \gaugesym, the observables change as follows. First, to facilitate the gauging, we need to Fourier transform the operators $\Phi_h(x;z)$ from the position $(x)$ to the momentum $(p)$ basis on the boundary. This gives rise to operators, which we will denote by $\Phi_h(p;z)$ (in a slight abuse of notation), which are eigenfunctions of the currents $(J^-,\bar J^-)$ with eigenvalues $(p,\bar p)$.  These operators behave like
\eqn\ppphhii{\Phi_h(p)= f_h(\phi)e^{i\vec p\cdot\vec\gamma}.}
Near the boundary at $\phi\to\infty$, one has $f_h(\phi)\sim e^{\beta\phi}$, with $\beta$  proportional to $h-1$.
Gauge invariance implies that the operators \locobs\ must be replaced in the deformed theory by
\eqn\observ{\hat\OO(p)=\int d^2z e^{-\varphi-\bar\varphi}\Phi_h(p)e^{-i(\omega t+p_yy)}\OO.}
The mass-shell condition \massshell\ is now deformed to
\eqn\mmaass{-{h(h-1)\over k}+{\alpha'\over4}(p_y^2-\omega^2)+\Delta_\OO=\half~.}
Moreover, gauge invariance sets $\omega=\epsilon p_0$ and $p_y=\epsilon p_1$. Observables corresponding to non-normalizable vertex operators, \observ\ with $\half<h\in\IR$, are labeled by two-dimensional momentum $(\omega, p_y)$, with $h$ fixed by the mass-shell condition \mmaass. One can use the coset description to calculate off-shell correlation functions of such observables, and use them to study the high (and low) energy behavior of the theory. Note that the observables \observ\ are labeled by their momenta. One does not expect to be able to Fourier transform them to position space, due to the non-locality of the theory. This is believed to be a general feature of all vacua of Little String Theory, such as DSLST \refs{\GiveonPX,\GiveonTQ,\AharonyXN}.

\newsec{Comments on the spectrum}

To study the spectrum of the theory, we would like to compactify the spatial direction on the boundary of the geometry \wwssll\ on a circle. In the undeformed theory (\ie\ for $\epsilon=0$), we can do this by identifying $\gamma_1\sim \gamma_1+2\pi R_1$, with all fields satisfying periodic boundary conditions on the circle. This gives rise to the $M=J=0$ BTZ black hole geometry, which describes a Ramond-Ramond ground state of the boundary CFT.

The spectrum of perturbative string states in this background is continuous. This is easy to understand from the spacetime point of view. The background \aaa\ is obtained by adding to a linear dilaton background, of the form $\IR_\phi\times \IR_t\times S^1\times \MM^4\times S^3$, $p$ fundamental strings wrapping the $S^1$ \GiveonZM. The resulting state is BPS -- the strings preserve some of the supersymmetry of the original background. Thus, these strings do not feel a force attracting them to the fivebranes, and their excitations form a continuum. This continuum is described by the vertex operators of long strings constructed in \refs{\MaldacenaHW,\ArgurioTB}.

The deformation \deltall\ extends the background from the near-horizon geometry of both the strings and the fivebranes to just that of the fivebranes. This extension does not change the fact that the strings experience a flat potential; hence, one expects to find a continuum of states corresponding to strings wound around the spatial circle in \wwssll, and having an arbitrary radial momentum in $\phi$ and oscillation level.

Such states can be described as follows using the coset description of the previous section. Consider for example the (NS,NS) sector vertex operators in eq. \observ.  To describe states carrying arbitrary momentum $n$ and winding $w$ around the $y$ circle, we replace the factor $e^{ip_yy}$ by
\eqn\plr{e^{ip_yy}\to e^{ip_L y_L+ip_Ry_R},}
with
\eqn\ppllrr{p_L={n\over R}+{wR\over\alpha'}\;;\;\;\; p_R={n\over R}-{wR\over\alpha'}~.}
States carrying real radial momentum correspond to
\eqn\formhhh{h=j+1=\half+is\;;\;\;\;s\in\IR.}
The mass-shell condition \mmaass\ now takes the form
\eqn\newmass{{\alpha'\over4}\omega^2={\alpha'\over4}p_L^2-{j(j+1)\over k}+\Delta_\OO-\half~,}
and a similar equation for the other worldsheet chirality,
\eqn\barnewmass{{\alpha'\over4}\omega^2={\alpha'\over4}p_R^2-{j(j+1)\over k}+\bar\Delta_\OO-\half~.}
Adding \newmass\ and \barnewmass\ gives
\eqn\finmass{\omega^2=\left(n\over R\right)^2+\left(wR\over\alpha'\right)^2+{2\over\alpha'}
\left(-{2j(j+1)\over k}+\Delta_\OO+\bar\Delta_\OO-1\right).
}
The difference of the two gives
\eqn\difftwo{\bar\Delta_\OO-\Delta_\OO=nw.}
For $w=1$, the mass-shell conditions \finmass, \difftwo\ describe a string winding once around the spatial circle on the boundary in a particular excitation state labeled by $\OO$ and with a particular radial momentum labeled by $s$ \formhhh. To rewrite it in a more suggestive form, it is useful to measure the energy of this state relative to the energy of a BPS string wrapping the circle (which corresponds to the supersymmetric vacuum), \ie\ write
\eqn\formomega{\omega=E+{R\over\alpha'}~.}
It is also useful to recall that in the $AdS_3$ limit of the background \wwssll, the last term in \finmass\ is related  to the value of $L_0$, $h_1$, for a long string with the same quantum numbers \refs{\MaldacenaHW,\ArgurioTB},
\eqn\recallh{\eqalign{h_1-{k\over 4}=&-{j(j+1)\over k}+\Delta_\OO-{1\over 2}~,\cr
\bar h_1-{k\over 4}=&-{j(j+1)\over k}+\bar\Delta_\OO-{1\over 2}~.}
}
Plugging \formomega, \recallh\ into \finmass, \difftwo, we find (for $w=1$) the mass-shell condition
\eqn\thuszamo{\left(E+{R\over\alpha'}\right)^2-\left({R\over\alpha'}\right)^2
={2\over\alpha'}\left(h_1+\bar h_1-{k\over 2}\right)+\left({n\over R}\right)^2,}
and $\bar h_1-h_1=n$. The mass-shell condition \thuszamo\ agrees precisely with what one would find for a state with the dimensions \recallh\ in a CFT $\MM$ of central charge $c_\MM=6k$, upon a deformation of the sort studied in \refs{\SmirnovLQW,\CavagliaODA}, $\delta\LL=-tT\bar T$. To determine $t$ it is useful to recall that the quantity $R$ in \refs{\SmirnovLQW,\CavagliaODA}, $R_{QFT}$, is in our language the circumference of the spatial coordinate on the boundary of $AdS_3$, $\gamma_1$, and is related to the circumeference of the $y$ coordinate at infinity, $2\pi R$, by a factor of $\epsilon$,{\foot{Due to the relation between $y$ and $\gamma_1$ mentioned above.} \ie\ $R_{QFT}=2\pi R\epsilon$. Similarly, the energy in these papers is related to the energy here by a factor of $1/\epsilon$. Taking all this into account, and comparing \thuszamo\ to the spectrum in \refs{\SmirnovLQW,\CavagliaODA}, we find $t=\pi\alpha'\epsilon^2$. 

As mentioned above, the physics is independent of $\epsilon$, since one can change it by rescaling the coordinates $(\gamma,\bar\gamma)$ (which also rescales the radius $R$) and shifting $\phi$. A convenient value  is $\epsilon=1$, since for that value the coordinates $(\gamma^0,\gamma^1)$ are normalized in the same way as the asymptotic coordinates $(t,y)$; this is the choice made in \GiveonNIE. Note that the value  $t=\pi\alpha'$, obtained here from the perturbative string spectrum (for $\epsilon=1$), agrees with that found in \GiveonNIE\ from black hole thermodynamics.

One can think of the states \thuszamo\ as belonging to the untwisted sector of the orbifold $\MM^p/S_p$. States with $w>1$ belong to the $Z_w$ twisted sector of the orbifold. To see this, one proceeds as follows. The analog of \formomega\ for this case is
\eqn\formomegaw{\omega=E+{wR\over\alpha'}~.}
Plugging this and \recallh\ into \finmass\ gives
\eqn\honehone{\left(E+{wR\over\alpha'}\right)^2-\left({wR\over\alpha'}\right)^2
={2\over\alpha'}\left(h_1+\bar h_1-{k\over 2}\right)+\left({n_w\over wR}\right)^2~,}
with $n_w=wn$. Comparing \honehone\ to \thuszamo, we see that the spectrum of strings with winding $w$ is the same as that of a string singly wound around a circle with radius $wR$ and momentum $n_w$. This agrees with the spectrum in the $Z_w$ twisted sector of $\MM^w/Z_w$, with $Z_w$ acting via cyclic permutation on the $w$ copies of $\MM$. Note that in the IR limit, $R/l_s\to\infty$, \honehone\ reduces to well known results in string theory on $AdS_3$ \refs{\MaldacenaHW,\ArgurioTB}, such as, \GiveonMI,
\eqn\hwhw{h_w={h_1\over w}+{k\over 4}\left(w-{1\over w}\right),}
describing long strings winding $w$ times around the boundary circle.

To recapitulate, we have shown that states with $w>0$ in the background \wwssll\ agree with those found in the symmetric product CFT $\MM^p/S_p$, with $\MM$ deformed via a $T\bar T$ deformation. Next we show that string theory on this background has states that do not fit this description, however these states decouple in the infrared limit, and thus are not visible in the IR CFT. 

Since states with winding $w$ correspond to the $Z_w$ twisted sector, it is natural to expect that states with $w\le 0$ are not captured by the symmetric orbifold. For $w=0$  \finmass\ takes the form
\eqn\wzero{(ER)^2={2R^2\over\alpha'}\left(-{2j(j+1)\over k}+\Delta_\OO+\bar\Delta_\OO-1\right),
}
where we also set the momentum $n=0$ for simplicity. Thus, the dimensionless energy $ER$ diverges in the limit $R^2/\alpha'\to\infty$, which means that the states \wzero\ are not present in the undeformed CFT dual to string theory on $AdS_3$. In the language of the $T\bar T$ deformed theory, these are states with energy $E\sim1/\sqrt t$, which decouple in the IR limit $t\to 0$. States with $w<0$ decouple even faster when $t\to 0$, as their energy is bounded from below by ${2 |w| R \over \alpha'}$.

\newsec{Discussion}

In the previous section we described perturbative string excitations of the system of strings and fivebranes wrapping the circle labeled by $y$ (or $\gamma_1$) in \wwssll, with the fivebranes wrapping an additional four dimensional surface $\MM^4$ \twelved. We saw that the excitations of this system that correspond to one or more of the $p$ strings creating the background moving away from the fivebranes (while remaining in their near-horizon region) are well described by a dual boundary theory $\MM^p/S_p$, where $\MM$ is a CFT with central charge $c_\MM=6k$, which roughly corresponds to the theory of a single string. The worldsheet deformation \deltall, which corresponds to $\epsilon\not=0$ in \wwssll, is dual to a $T\bar T$ deformation (in the sense of \refs{\SmirnovLQW,\CavagliaODA}) of the CFT $\MM$. 

There are also states which do not fit the $\MM^p/S_p$ structure, but they do not correspond to small excitations of the string/fivebrane system. States with $w<0$ can be thought of as obtained from the system of $p-1$ strings and $k$ fivebranes by adding to it an $F1-\bar{F1}$ pair, while states with $w=0$ correspond to adding to the string/fivebrane system an additional short string. In contrast, states with $w>0$ can be thought of as describing excitations of the $p$ strings forming the vacuum. 

In this note, we focused on the spectrum of excitations of a particular Ramond-Ramond vacuum of the spacetime CFT, corresponding to the $M=J=0$ BTZ black hole. Modular invariance, spectral flow symmetry of the spacetime CFT, and explicit constructions imply the existence of many other Ramond and Neveu Schwartz vacua, and it would be interesting to extend our discussion to these vacua. We will postpone this to another publication. 

Our results are related to those of \GiveonMI, which discussed the transition between perturbative strings and black holes in $AdS_3$ and linear dilaton backgrounds. The picture presented in that paper was the following. As one increases the available energy, first one creates perturbative long strings that can propagate towards the boundary. As the energy of these strings increases, they propagate to larger and larger radial distance, where the coupling of the theory on the strings grows, and eventually they cross over to black holes. In this sense, one can think of the long  strings as precursors of the black holes. Our results reinforce and extend this picture. The agreement found in  \GiveonNIE\ and here suggests that the symmetric product provides a good description of both the long strings and the black holes in $AdS_3$, and the deformation \deltall\ corresponds to the $T\bar T$ deformation in $\MM$.

Note that the above discussion is valid for the case where the level of the $SL(2,\IR)$ current algebra $k$ is larger than one. As discussed in detail in \GiveonMI, for $k<1$ the physics is different -- the black holes discussed in \GiveonMI\ are not normalizable, the coupling on the long strings in $AdS_3$ becomes weak near the boundary, and the generic high-energy states are these long strings. All this agrees with the $T\bar T$ deformed symmetric product theory, where in this case the spacetime CFT on $\MM$ is one in which the $SL(2,\IR)$ invariant vacuum is not in the spectrum.

\bigskip\bigskip
\noindent{\bf Acknowledgements:}
We thank O. Aharony, J. Maldacena, E. Martinec, N. Warner and E. Witten for discussions.
The work of AG and NI is supported in part by the I-CORE Program of the Planning and Budgeting Committee and the Israel Science Foundation (Center No. 1937/12), and by a center of excellence supported by the Israel Science Foundation (grant number 1989/14). DK is supported in part by DOE grant DE-SC0009924. DK thanks Tel Aviv University and the Hebrew University for hospitality during part of this work.

\listrefs
\end

and the last term in \finmass\ is thus related  to the value of $L_0$, $h_w$,
for $w$ long strings with the same quantum numbers, by
\refs{\MaldacenaHW,\ArgurioTB}
\eqn\recallhw{\eqalign{w\left(h_w-{kw\over 4}\right)=&-{j(j+1)\over k}+\Delta_\OO-{1\over 2}~,\cr
w\left(\bar h_w-{kw\over 4}\right)=&-{j(j+1)\over k}+\bar\Delta_\OO-{1\over 2}~.}
}
Plugging \formomegaw,\recallhw\ into \finmass,\difftwo, we find the mass-shell condition
\eqn\thuszamow{\left(E_w+{wR\over\alpha'}\right)^2-\left({wR\over\alpha'}\right)^2
={2w\over\alpha'}\left(h_w+\bar h_w-{kw\over 2}\right)+\left({n\over R}\right)^2,}
and $\bar h_w-h_w=n$. The mass-shell condition \thuszamow\ agrees apparently with what one would find for a state with the scaling dimensions $h_w-{c_w\over 24}$, $\bar h_w-{c_w\over 24}$, in a CFT of central charge $c_w=6kw$, upon a $t_wT\bar T$ deformation of the sort studied in \refs{\SmirnovLQW,\CavagliaODA}.
Moreover, \thuszamow\ can be rewritten as
\eqn\zamoww{\left({E}+{R\over\alpha'}\right)^2-\left({R\over\alpha'}\right)^2
={2\over\alpha'}\left({h}+{\bar h}-{k\over 2}\right)+\left({n/w\over R}\right)^2,}
where $\{E,h,\bar h\}=\{E_w,h_w,\bar h_w\}/w$,
which agrees precisely with what one would find for a state with the scaling dimensions
$h-{c_\MM\over 24}$, $\bar h-{c_\MM\over 24}$, in a CFT $\MM$ of central charge $c_\MM=6k$,
upon a $tT\bar T$ deformation to a state with energy ${E}$ and momentum ${n/w}$, of the sort studied in \refs{\SmirnovLQW,\CavagliaODA},
with a deformation parameter $t=\pi\alpha'$, that is identical to the one in (6.12) of \GiveonNIE,
and which is $w$ times bigger than the apparent one, $t_w={\pi\alpha'\over w}$, in \thuszamow.
Equations \thuszamow\ and \zamoww\ are compatible with the expectation that states in the $w$ winding sector
of the perturbative string spectrum around the $M=J=0$ BTZ black hole background,
correspond in the spacetime theory to states such as in the $w$-twisted sector of a symmetric orbifold $\MM^p/S_p$, with the deformation \deltall\ acting on the CFT $\MM$ as in \refs{\SmirnovLQW,\CavagliaODA}.

Comment:

Recall \GiveonMI\ that \recallhw\ implies that
\eqn\hwhw{h_w={h_1\over w}+{k\over 4}\left(w-{1\over w}\right)~.}
This is precisely the type of expression one finds in the $w$-twisted sector of a symmetric
orbifold, whose building block $\MM$ has $c_\MM=6k$.
In particular, plugging \hwhw\ into \thuszamow,
one finds
\eqn\honehone{\left(E_w+{wR\over\alpha'}\right)^2-\left({wR\over\alpha'}\right)^2
={2\over\alpha'}\left(h_1+\bar h_1-{k\over 2}\right)+\left({n_w\over wR}\right)^2~,}
which is the same as the equation for the $w=1$ sector, \thuszamo, with $R\to wR$, an energy $E_w$
above extremality, and momentum $n_w=wn$,
in harmony with a symmetric orbifold structure.